# Effect of Bend Loss on Parabolic Pulse Formation by Active Dispersion Tailored Fibers

Dipankar Ghosh and Mousumi Basu

*Abstract*— **This work reports the performances of straight and bent active normal dispersion decreasing fibers (NDDF), with spatial nonlinear variation, to form parabolic self-similar pulses. The core radius changes along the NDDF length, thereby altering the transverse field distribution lengthwise. Hence bend loss is no longer a constant quantity. Including this loss variation, we investigate the performances of NDDFs as generators of parabolic self-similar pulses. In view of the small changes of relative refractive index differences (Δ) during production, we obtain several NDDFs with variations of core radii. Suitable choice of Δ and bend radius of curvature of the fibers leads to obtain similaritons. Even for sufficiently small Δ and bend radii, parabolic pulses are formed at the cost of higher optimum length in comparison to straight fibers. The comparative study on the straight and bent NDDFs with different Δ values is helpful for fiber optic manufacturers to fabricate the proposed NDDFs.**

*Index Terms*— **Active Normal dispersion decreasing fiber, Bend loss, nonlinear Schrödinger equation, self-similar parabolic pulse.**

PARABOLIC similaritons with a linear chirp and less wave breaking are very much important in the context of present day research [1-3] in high power lasers, pulse compression or supercontinuum generation. Stable similariton pulses are created by hyperbolic dispersion variation [3] along the length of a normal dispersion decreasing fiber (NDDF). Similariton regime starts at relatively larger length for passive NDDFs in comparison to the active fibers. The combination of dispersion tapering with gain [4-5] is also reported as useful for designing appropriate devices. In this work the active NDDFs are studied to create self-similar parabolic pulses, by including the spatial nonlinear variations [6] as well as the external gain. As fiber bends are unavoidable for such fibers which have mainly device applications, the effect of bend loss [7] on the NDDFs are investigated for production of parabolic similaritons.

The pulse propagation in active NDDFs in presence of an external gain can be expressed by the nonlinear Schrödinger equation (NLSE) [4-6,8], which after some steps of appropriate coordinate transformation [6], can be written as

$$i\frac{\partial u}{\partial \zeta} - \frac{1}{2}\frac{\partial^2 u}{\partial \tau^2} + N^2|u|^2 u = i\frac{\delta_{eff}}{2}u, \qquad (1)$$

This work was supported by Council of Scientific & Industrial Research, Govt. of India, through Senior Research Fellowship to D. Ghosh.

Authors are with Department of Physics, Bengal Engineering and Science University, Shibpur, Howrah – 711103, West Bengal, India (e-mail: dippghosh@yahoo.com; mousumi_basu@yahoo.com).

where $N^2 = \gamma(0)P_0 T_0^2/\beta_2(0)$, $\beta_2(0)$ and $\gamma(0)$ are the initial values of group velocity dispersion (GVD) factor and nonlinear factor, $P_0$ and $T_0$ are the initial peak power and pulse width and $\tau$ is the normalized time [5]. Here $u$ and $\zeta$ represent the transformed pulse envelope and normalized coordinate respectively, dependent on the longitudinal variations of normalized dispersion and nonlinearity as given by [6]. The effective gain coefficient ($\delta_{eff}$) in (1) is governed by the virtual gain induced by the dispersion variation {$D(z)$} as well as a nonlinear variation {$\Gamma(z)$}[6], in addition to the physical gain coefficient {$\delta_0 = GL_d$}. $L_d$ is the dispersive length and $G$ represents the external gain over the attenuation which is considered to be 8 km⁻¹ here. As in presence of bending [7], the fibers also exhibit a longitudinal loss variation and thus the effective gain is governed by

$$\delta_{eff}(z) = \frac{1}{D(z)}\left\{\delta_0 - \alpha_b(z)L_d - \frac{1}{D(z)}\frac{dD(z)}{dz} + \frac{1}{\Gamma(z)}\frac{d\Gamma(z)}{dz}\right\}. \qquad (2)$$

To design a NDDF with parabolic index profile, we choose a GeO₂ doped core dispersion compensating fiber (DCF) with Δ = 0.84% and initial GVD factor $\beta_2$ (0) = 12.75 ps²/km. The GVD variation (shown in Fig.1) is considered from the normalized variation of $D(z)$, given by Wabnitz et al. [5]

$$D(z) = \frac{\delta_0}{\delta}\left[1 + \frac{\delta - \delta_0}{\delta_0 + \delta\{\exp(\delta_0 z/L_d) - 1\}}\right], \qquad (3)$$

where $\delta$ represents the gain coefficient (taken as 1.05), dependent on physical gain coefficient $\delta_0(\sim 0.63)$ as well as the virtual gain induced by $D(z)$. Generally the dispersion profile can be easily obtained from the lengthwise variation of core radius ($a$). However, there is no simple method to obtain the variation of '$a$' with '$z$' from previously known dispersion profile. Developing a series of programs we found out the appropriate forms of $a(z)$ from $D(z)$. If there is a slight change ($\sim 10^{-4}$) of Δ during production, $a(z)$ will be different in each case, producing changes in modal spot sizes {$w(z)$} [8-9], which in turn alters the nonlinear factor {$\gamma(z)$} of the fiber. As $\gamma$ is inversely proportional to $w^2$ [9], Fig.1 shows that $\gamma$ increases from initial value of $\gamma(0) \sim 1.95$ W⁻¹km⁻¹ and saturates to ~2.9 W⁻¹km⁻¹ near 500 m with Δ = 0.0084. Similar natures are followed for other NDDFs with ±1.2% of Δ variations (0.0083-0.0085). As the modal field distribution along the length changes, the bend loss ($\alpha_b$ in dB/length) of NDDFs can not be avoided. Here we estimate $\alpha_b$ for different bend radii of

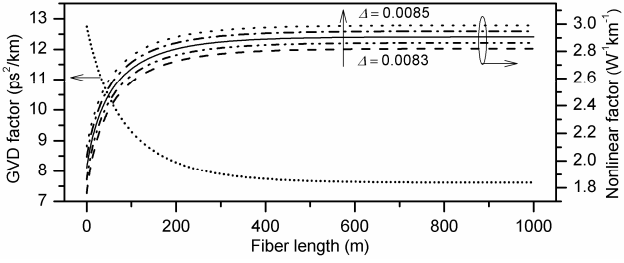

Fig. 1. Lengthwise variations of GVD factor (left) and nonlinear factors (right) for a range of relative refractive index differences (Δ) (0.0083-0.0085) of the NDDF.

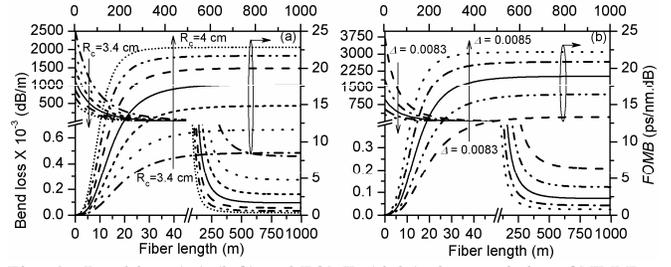

Fig. 2. Bend loss ($\alpha_b$) (left) and FOMB (right) characteristics of NDDFs as a function of fiber length (z) (a) for different $R_c$ with Δ = 0.0084 and (b) for different Δ with $R_c$ = 3.75 cm.

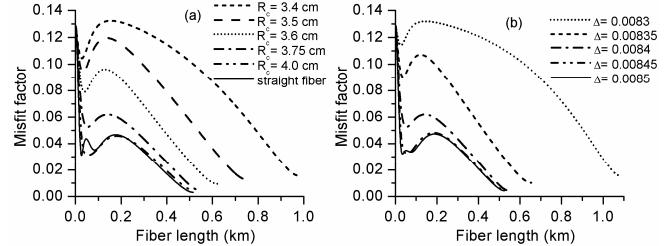

Fig. 3. Variation of misfit parameter with fiber length for (a) bent fibers with different radii of curvatures as well as for straight fiber when Δ = 0.0084 and (b) bent fibers with $R_c$ = 3.75 cm for different Δ values.

curvatures ($R_c$) of the NDDFs with the following relation [7],

$$\alpha_b = 4.343\sqrt{\pi V^8 / 16 a R_c W^3} \, exp\left\{-\left(4 R_c \Delta W^3\right)/\left(3 a V^2\right)\right\} \times I \quad (4)$$

where $I = \left[\int_0^\infty \left\{1-f(R)\right\} F_0(R) I_0(WR) \, R \, dR\right]^2 / \int_0^\infty F_0^2(R) R \, dR \quad (5)$

and $W = V\sqrt{b}$, the normalized propagation constants ($b$) are measured as $b = \left(n_{eff}^2 - n_2^2\right)/\left(n_1^2 - n_2^2\right)$ at $V$-parameters [9], $n_{eff}$ being effective index of the medium, $n_1$ and $n_2$ are the central core and clad index. $I_0(WR)$ is the modified Bessel function, $R$ being the normalized radial coordinate. $f(R)$ is the profile shape ($R^2$ here) and $F_0(R)$ represents the transverse field distribution, from which '$w$' is obtained. When '$w$' changes with $z$, we get '$\alpha_b$' as a function of fiber length from (4). Incorporating this '$\alpha_b$' along with other fiber losses ($\alpha$), the figure of merit $\left\{FOMB = D_c/(\alpha_b + \alpha)\right\}$ in ps/nm.dB [10] along the bent NDDF length is estimated. The '$\alpha_b$' characteristics of the NDDF with Δ = 0.84% is shown in Fig. 2(a). For larger $R_c$ (4 cm), the smallest $\alpha_b$ profile is seen with relatively larger FOMB, compared to fibers with other $R_c$ values. Slight decrease in $R_c$ is followed by adequate decrease of FOMBs. In Fig. 2(b), $\alpha_b(z)$ and $FOMB(z)$ are shown, for a range of Δ values. The loss is too large (~ 3700 dB/km) for smaller Δ = 0.83% at $z$ = 0 and approaches to ~ 0.21 dB/km at the output of 1 km long NDDF with $FOMB \sim 13.3$ ps/nm.dB. For larger Δ (~ 0.85%), $\alpha_b$ and $FOMB$ saturate to ~ 0.028 dB/km and ~ 22.2 ps/nm.dB respectively after 500 m of propagation.

To study the generation of self-similar parabolic pulses from the active NDDFs, we solve the NLSE by Symmetrized Split Step Fourier method [8], with an initial Gaussian pulse of energy 88.6 pJ. The calculated misfit parameters ($\mu$) [6] of output pulses for straight and bent NDDFs with their longitudinal variations in Fig. 3, confirm the formation of parabolic similaritons, when $\gamma$ variations are included. In Fig. 3(a) the solid curve shows that $\mu$ is ~ 0.003 at the optimum length ($z_{opt}$) of 515 m for similariton formation in a straight fiber, whereas $z_{opt}$ increases appreciably for bent fibers. For adequate bending ($R_c$ = 3.4 cm), $\mu$ is ~ 0.01 at $z_{opt}$ = 982 m. We also show that for CCITT recommended $R_c$ = 3.75 cm, lowest $\mu$ (~ 0.004) is achieved at $z_{opt}$ = 542 m, which is 27 m more than that of a straight fiber. The misfit characteristics of bent fibers with $R_c$ = 3.75 cm, for a range of Δ (0.83-0.85%) are shown in Fig. 3(b). Decrease in Δ by 1.2%, (0.0084 to 0.0083) makes $z_{opt}$ too high, ~1.087 km, whereas increase in Δ to 0.0085 reduces '$\alpha_b$' and as a consequence $z_{opt}$ is also reduced to 526 m. In Fig. 4, the contour plots of intensity profiles show the temporal broadening of the pulses as well as the increase in intensity for the NDDF with different $R_c$ values. Comparing Fig. 4(a) with Fig. 4(f), it is seen that the output intensity for a bent fiber ($R_c$ = 3.4 cm) becomes much less (~ 38.7 W) at a large $z_{opt}$ ~ 982 m than that of a straight fiber for which output intensity is more than 110 W at adequately smaller $z_{opt}$ of 515 m. As '$\alpha_b$' for NDDF with $R_c$ = 3.4 cm is too high in the first 100 m, the system loss dominates in spite of the presence of $\delta_0$. Thus we see no trace of pulse intensity until '$\alpha_b$' starts to decrease sufficiently. After that the intensity gradually increases with distance along with an increase in the effective pulse width. The contour plot in Fig. 4(e) shows that increase in $R_c$ to 4 cm, leads to obtain amply high output power (~ 63.8 W) at $z_{opt}$ ~ 516 m, which is very close to the $z_{opt}$ for straight fiber. Here the effective gain of the fiber is able to play a leading role over the bend loss of the system after a small distance, in comparison to the cases in Fig. 4(a)-(d), although the output power is slightly low compared to that of a straight NDDF. Figure 5 depicts the reduction of parabolic output pulse intensity at $z_{opt}$ due to increase in loss, when $R_c$ is decreased up to 3.5 cm. For $R_c$ = 3.4 cm, the power becomes slightly high due to the effect of gain over a long distance. Here we obtain almost constant spectral power over a frequency range $\Delta v$ ~ 2.0 THz for the straight NDDF. A noticeable decrease in spectral width having constant spectral power is seen for sufficiently bent fibers. Considering the changes in Δ (0.83-0.85%), the output power as well as the rms pulse widths ($\sigma_{rms}$) [8] and spectral widths ($\Delta v_{rms}$) [8] are estimated and plotted as a function of bent NDDF ($R_c$ = 3.75 cm) length ($z$) in Fig. 6(a) and (b) respectively. It is seen when Δ = 0.85%, $\sigma_{rms}$ and $\Delta v_{rms}$ increases to ~ 13.7 ps and 0.71 THz, respectively at $z_{opt}$ = 526 m, whereas both rms values are much less (~ 8.5 ps and 0.33 THz respectively) when $z_{opt}$ is nearly doubled to 1.09 km for a lower $\Delta$ = 0.83%. The plot of $z_{opt}$ for obtaining similaritons, as a function of $R_c$ for the NDDF profile with $\Delta$ = 0.0084 in Fig.7, shows that initially $z_{opt}$ decreases sharply from a very large value because of the exponential dependence of '$\alpha_b$' on $R_c$ and gradually approaches to the value of $z_{opt}$ for a straight fiber. The variation of $z_{opt}$ with $\Delta$, illustrates a perfectly linear nature for straight fibers. For a bent fiber ($R_c$ = 3.75 cm), when $\Delta$ is increased, $z_{opt}$ (left side) reduces much and finally approaches towards the $z_{opt}$ for straight fiber. Thus the study implies that the bend insensitive nature of the proposed NDDF can be optimized for relatively higher $\Delta$ values.

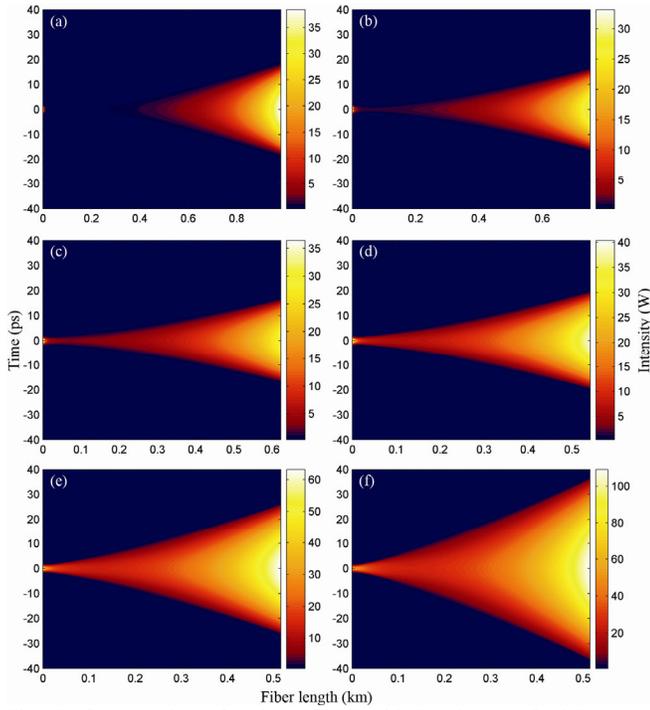

Fig. 4. Contour plots of (a) bent NDDF with $R_c$ = 3.4 cm, (b) 3.5 cm, (c) 3.6 cm, (d) 3.75 cm, (e) 4.0 cm and also of (f) straight NDDF having $\Delta$ = 0.0084.

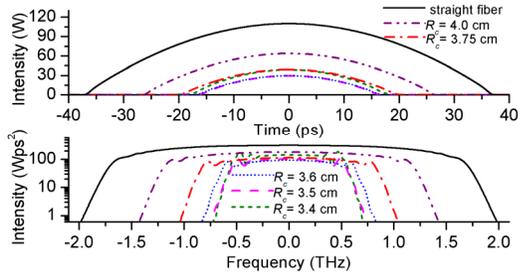

Fig. 5. Output temporal (top) and spectral (bottom) intensity profiles of different bent and straight NDDFs having $\Delta$ = 0.0084.

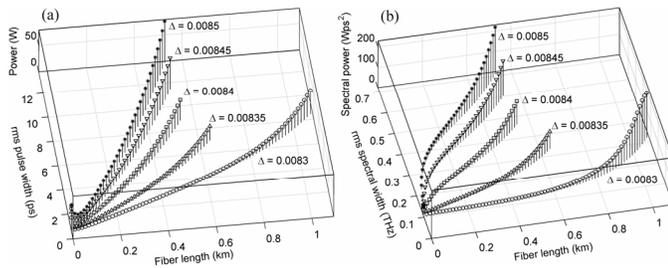

Fig. 6. Power and rms width variation with length of bent fibers ($R_c$ = 3.75 cm) for different $\Delta$ values in (a) temporal and (b) spectral domain.

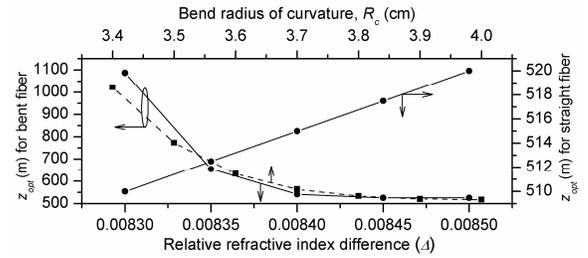

Fig. 7. Dependence of optimum fiber length ($z_{opt}$) on $R_c$ for given $\Delta$ = 0.0084 and on $\Delta$ for given $R_c$ = 3.75 cm.

In conclusion, we successfully presented the effect of bending on the performance characteristics of a designed active NDDF for generation of parabolic pulses, when the nonlinear variation along length is taken into account. As the transverse field distribution changes along length, bend loss is no longer an invariable quantity throughout the propagation length and we only incorporated this bend loss variation to study the similariton generation in the NDDF. In view of the changes in $\Delta$ during fabrication, the bent fiber characteristics show that bend insensitive NDDFs with optimized $\Delta$ values would be proper choices for similariton generation. Even for highly bent fibers with smaller $\Delta$ values, parabolic pulses can be generated at the cost of higher optimum lengths, compared to that of a straight fiber. The studies on the variations of optimum length as a function of $R_c$ and $\Delta$ values are helpful for fiber manufacturers to fabricate active NDDFs, competent enough to produce similaritons even in presence of bending.